**Modeling Epidemics on adaptively evolving networks: a data-mining perspective**

Assimakis A. Kattis[a], Alexander Holiday[a], Ana-Andreea Stoica[b] & Ioannis G. Kevrekidis[a,c]

[a] Department of Chemical and Biological Engineering, Princeton University, Princeton, NJ 08544, USA

[b] Department of Mathematics, Princeton University, Princeton, NJ 08544, USA

[c] Department of Chemical and Biological Engineering and Program in Applied & Computational Mathematics, Princeton University, Princeton, NJ 08544, USA

**Abstract**

The exploration of epidemic dynamics on dynamically evolving ("adaptive") networks poses nontrivial challenges to the modeler, such as the determination of a small number of informative statistics of the detailed network state (that is, a few "good observables") that usefully summarize the overall (macroscopic, systems level) behavior. Trying to obtain reduced, small size accurate models in terms of these few statistical observables –that is, coarse-graining the full network epidemic model to a small but useful macroscopic one- is even more daunting. Here we describe a *data-based approach* to solving the first challenge: the detection of a few informative collective observables of the detailed epidemic dynamics. This will be accomplished through Diffusion Maps, a recently developed data-mining technique. We illustrate the approach through simulations of a simple mathematical model of epidemics on a network: a model known to exhibit complex temporal dynamics. We will discuss potential extensions of the approach, as well as possible shortcomings.

**Introduction**

Mathematical modeling of epidemic dynamics is an indispensable tool in understanding and mitigating the spreading of disease in the real world.[1-4] As computational power, numerical simulation techniques

and –most recently- "big data" tools and techniques progress, the degree of realism in these mathematical models constantly improves. From simple nonlinear models of Susceptible-Infected-Recovered (SIR) or Susceptible-Infected-Susceptible (SIS) dynamics that are based on spatial averaging (so called mean-field models, consisting of a few nonlinear ordinary differential equations (ODEs)) we have graduated to models with detailed spatial information and structure, incorporating not only geographical details of communities and cities but also information about the social interactions between the individuals involved.[5-7] From mean field ODEs the models now become large-scale, stochastic, individual-based simulations on networks (geographical as well as social).  And while this framework is convenient for investigating many different initial conditions combined with many different network connectivities and many different interaction/evolution rules, recording and rationalizing a useful summary of the dynamics (that is, the relevant macroscopic, systems-level *statistics* of these scenarios) is crucial for systems-level understanding and control. Finding the right macroscopic observables for such detailed simulations- the right variables to summarize the epidemic dynamics is still a daunting task.

Over the last decade our group has proposed –and developed- the so-called Equation-Free computational framework for complex/multiscale systems modeling: given a detailed (here, individual/agent-based) simulation algorithm, this framework enables us  to study coarse-grained, systems level dynamics through the design, execution and processing of the brief bursts of fine scale simulation data; Equation-Free algorithms like Coarse Projective Integration (CPI) take the form of  "wrappers" around the fine scale code (say, an agent-based epidemic simulation code on an adaptive network).[8-11] Yet for this approach to be successful, one needs to *a priori* know *what the right macroscopic statistics are* (e.g., the right few leading moments of the distribution of susceptible or of infected individuals in the population) in terms of which the epidemic statistics can be informatively  summarized.

This paper considers the case where such informative and parsimonious system-level statistics are not *a priori known.*  In this case the equation-free modeling approach can still be carried through, as long as the right macroscopic variables can be discovered *through the mining of (big) computational simulation data.* This is a "doubly-data-based" modeling strategy: using data produced from detailed, individual level, fine scale simulation bursts *to detect the number and identity of the macroscopic observables;* and then, armed with this knowledge, design and execute new, informative, microscale simulations to systematically explore the evolution of the epidemic. This jointly "equation-free, variable-free" approach holds great promise for the accurate, fast and informative systems-level simulation of detailed, realistic epidemics models – the "right observables" come from the (previously computed) data, and the design of "the right simulations" to obtain useful new information also comes from the (previously computed) data.

When the state of the fine-scale model at a given moment in time can be mathematically described as a (long) vector in $R^n$ (e.g. the state of a large number $n$ of agents), both linear data-mining techniques, such as Principal Component Analysis (a technique based on the Singular Value Decomposition –SVD- of data matrices), as well as nonlinear data-mining techniques such as ISOMAP or Diffusion Maps (DMAPS) can be applied to simulation data ensembles to obtain "the right" macroscopic observables.[12-14] But when the data involve *evolving graphs* (and in our case, we are interested in epidemic dynamics on adaptively evolving networks), finding good macroscopic observables based on simulation databases is a nontrivial task. We will present here a simple modification/extension of the DMAPS procedure that will allow us to detect the (small) number of these observables for epidemics on adaptive networks based on data mining only. The "macro-variables" discovered by this process for the case of a SIS epidemic on an adaptive network will be presented, discussed, and contrasted to traditionally used macro-variables for the same problem.

Our approach is motivated by, and illustrated through, SIS dynamics on adaptive networks.[1,15,16] The computational methodology, however, is in principle applicable to many problems that involve the dynamic evolution of networks with both labeled nodes (when we know the identity of the individuals) or unlabeled nodes (when we do not). [17,18] We will link the DMAPS procedure with quantities that allow us to usefully compare different networks; we will use this approach to detect the number of macroscopic observables involved in the dynamics of our SIS epidemic model, and compare these data-based observables with typical network statistics. The last few years have seen several innovative approaches to finding accurate reduced models for dynamic, network-evolution problems, extending and complementing well-established techniques like those based on moment closures. Our approach should be considered as a data-mining based alternative to these techniques.

The rest of the paper is organized as follows: We will first describe our implementation of a SIS epidemic model on an adaptive network, from which the simulation data will be obtained. We will then briefly discuss established data mining techniques: a linear technique (PCA) and a nonlinear one (Diffusion Maps). We will then present the extension to Diffusion Maps that expands their applicability to data in the form of evolving networks (where the connectivity of the network evolves in time along with the state of the network nodes). We first validate our network data mining approach on data obtained from a simple Watts-Strogatz network model.[19] We then present our main results: the application of our data-mining technique to data collected from dynamic SIS simulations on an adaptive network, obtained over a range of epidemic parameter values where it is known that complex, oscillatory dynamics arise. We discuss the relation of the variables detected through our approach to those of more traditional, moment-based approaches, and conclude with a brief perspective on potential shortcomings but also potential fruitful applications of the approach.

**The Adaptive SIS Model**

An implementation of the adaptive SIS model can be constructed by considering a labeled graph $G$ with $N$ nodes and $L$ links, with each node representing an individual in a social network; the state of each individual, either susceptible (S) or infected (I), constitutes the label of the node. Edges between individuals are defined as SS-links, II-links, or SI-links, according to the label of the nodes they connect. Starting with a given initial network connectivity pattern (a given network topology), the evolution of the model can be characterized by three substeps, which together constitute a time step in the model's evolution:

1. All infected nodes recover with probability $r$, becoming susceptible.
2. For every SI-link, the susceptible individual becomes infected with probability $p$.
3. Every SI-link is removed with probability $w = w_0\rho$. In this case, a new edge between the corresponding susceptible node and another susceptible node is formed. The new link is made with a node chosen uniformly at random from the set of all other susceptible individuals.

The probability of rewiring $w = w_0\rho$ is based on a constant input parameter $w_0$ and the infected fraction $\rho = i/N$, where $i$ is the number of infected nodes. These rules are motivated by the assumption that humans are more likely to avoid infected individuals proportionally to their awareness of disease spread, which here is assumed to be directly proportional to the infected fraction of the population.[1]

Although the behavior of this model is inherently complex (see the bifurcation diagram in Fig. 1A, reproduced by permission),[16] it has been previously established that the system-level dynamics of a sufficiently large network can be captured by just three macroscopic observables: the number of infected nodes $i$, the number of SS-links $l_{SS}$ and the number of II-links $l_{II}$. These variables suffice to describe all long-term dynamical behavior exhibited by the system, since the values of other variables (higher order moments of the network state) quickly become slaved to (quickly become functions of) these three and do not contribute extra degrees of freedom over long timescales. As the infection parameter $p$ varies, one can observe stationary states as well as coarsely oscillatory dynamics, and even coexistence between the two (associated with an apparent subcritical coarse Hopf bifurcation).[20] In this paper we will show how to extract the relevant observables responsible for the dynamics of the system *without* making use of any prior knowledge about which ones are most suitable. This will be accomplished by first constructing a suitable similarity measure for quantifying graph differences, and secondly, by applying Diffusion Maps on ensembles of graphs resulting from the dynamic simulation of the model's evolution.

**A Brief Discussion of Dimensionality Reduction**

Analysis of the dynamics of this epidemic model, especially as the size of the network grows, is hindered by its size and complexity. Not only are there many nodes to keep track of over thousands of time steps, but each step is also comprised of a number of different (stochastic) actions. These factors combine to make *systematic* exploration of such a system computationally intractable; one may only execute and observe many different scenarios computationally (different initial networks, different initial node states, different parameter values). Additionally, it is not clear which variables, or indeed even how many, play a determining role in summarizing long-term system behavior. This motivates the development of an algorithmic approach to identifying the crucial features of the system. Not only will these important features themselves aid in better understanding the problem and in summarizing its behavior, but they could also be used in an equation-free framework to enable the sort of analysis typically reserved for simpler systems (e.g. a few mean-field ODEs). Below, we present our first step towards this goal: the use of the DMAPS data mining technique to identify the important observables in the SIS model from simulation datasets.

**Principal Component Analysis and Diffusion Maps**

Given a dataset $\{x_1, x_2, \ldots, x_N\}$ (with each $x_i \in \mathbb{R}^n$) several approaches exist to uncover a simpler (coarse-grained, reduced, system-level) description $\{y_1, y_2, \ldots, y_N\}$ of the data, where $y_i \in \mathbb{R}^p$ and $p \ll n$. These new lower-dimensional points $y_i$ accurately capture the salient features of the original data set, with minimal loss of important information. In essence, we try to describe our data as concisely as possible.

Perhaps the best-known method for achieving this is Principal Component Analysis (PCA).[12] This method attempts to capture the important features of the data by identifying linear correlations among data points. Thus, if the three-dimensional input points $x_i \in \mathbb{R}^3$ actually only span a two-dimensional plane, PCA would discover two basis vectors spanning the plane itself, allowing each of the original three-dimensional points to be efficiently described with just two new coordinates (a process illustrated in Fig. 3). Unfortunately, this technique assumes the data lies on, or around, a linear subspace, whereas data points will, in general, lie on nonlinear manifolds. To circumvent this limitation we turn to the nonlinear dimensionality reduction technique of Diffusion Maps (DMAPS).[13] The (somewhat technical) algorithm is outlined below.

Given $N$ vectors $\{x_1, x_2, \ldots, x_N\}$ in $\mathbb{R}^n$, we form an $N \times N$ matrix $W$ defined as:

$$W_{ij} = \exp\left(-\frac{d(x_i, x_j)^2}{\epsilon^2}\right),$$

where $d(x_i, x_j)$ is a measure of the distance between points $x_i$ and $x_j$, (usually taken as the Euclidean distance), and $\epsilon$ represents the neighborhood in which we consider $d(x_i, x_j)$ to be "a meaningful metric". By defining the diagonal matrix $D_{ii} = \sum_j W_{ij}$, we can create a new row-stochastic matrix:

$$A = D^{-1}W.$$

The entries of $A$ can be viewed as defining a random walk over the dataset. The first few eigenvectors of this random walk process represent an efficient parameterization of the original high-dimensional dataset. We denote the *k-dimensional diffusion map embedding* at time $t$ as the transformation:

$$\Phi_t^{(k)}(x) = (\lambda_2^t \phi_2, \lambda_3^t \phi_3, \ldots, \lambda_{k+1}^t \phi_{k+1}),$$

where $\lambda_i$ and $\phi_i$ are the i[th] eigenvalue and eigenvector respectively, ordered by decreasing magnitude; throughout this paper we set $t = 0$.

By simulating a diffusion process over the dataset, DMAPS will reveal the underlying low-dimensional nonlinear structure. This is illustrated in Fig. 3, where the algorithm is applied to a collection of points $\{x_1, x_2, \ldots, x_N\}$, $x_i \in \mathbb{R}^3$ that lie on a curved surface (a "Swiss roll"). The result, also shown in Fig. 3 is a concise, two-dimensional embedding of the data into $\{y_1, y_2, \ldots, y_N\}$, $y_i \in \mathbb{R}^2$, giving a more efficient description of each point by its coordinates along the length and width of the underlying plane. Note that PCA would not produce meaningful two-dimensional results in this context, as the relations between input variables are nonlinear.

**Similarities between Different Networks**

In order to simulate diffusion over the dataset, DMAPS requires a scalar distance between two data points, $d(x_i, x_j)$. When each point $x_i$ is a vector in $\mathbb{R}^n$ as in our examples in Fig. (3), the Euclidean distance between points is often sufficient. However, in the dataset we investigate below, each point $x_i$ is not a vector, but actually a graph (or network), which we represent as $G_i$.

The literature contains a number of ways of quantifying the distance between two graphs, $d(G_i, G_j)$, for example by measuring how easily $G_i$ can be transformed into $G_j$, the Graph Edit Distance, or by comparing random walks on the graphs (a spectral distance).[21-24] In this paper, we consider two graphs to be similar if they share similar numbers of certain features. More precisely, we define a list of $k$ subgraphs $S = \{s_1, s_2, \dots, s_k\}$, such as the single edge, the two connected edges, the triangle shown in Fig. (4) etc. Then we record how many times each subgraph appears in our input graph in a vector $v_i = \{c_1^i, c_2^i, \dots, c_k^i\}$, where $c_j^i$ is the number of times subgraph $s_j$ was found in input graph $G_i$. This process maps each graph to a vector of subgraph densities, the counts $G_i \rightarrow v_i$, $v_i = \{c_1^i, c_2^i, \dots, c_k^i\}$ in $\mathbb{R}^k$, thus embedding the graph as a point in $\mathbb{R}^k$. We then use these k-long vectors (k-dimensional points) as the input to DMAPS, and use the Euclidean distance between them as our notion of graph similarity. Thus $d(G_i, G_j) = \|v_i - v_j\|_2$. Fig. (4) presents a schematic illustrating this subgraph-enumeration process.

In the SIS model, we are actually working with labeled graphs, since each node has one of two labels – susceptible (S) or infected (I). It is inappropriate to simply ignore network labels, since networks with the same connectivity, but with different node labels, can behave in extremely different ways and should thus be considered dissimilar. To overcome this issue, for a given labeled graph $G$ we choose to consider three separate unlabeled subgraphs $G_0, G_S$ and $G_I$. Here $G_0$ is the initial graph $G$ without node labels and $G_S, G_I$ the unlabeled subgraphs obtained (induced) by only considering the S, or only the I nodes respectively. We will represent each overall labeled graph by the concatenation of the three count vectors $v_0, v_I, v_S$ into $v = [v_0 \ v_I \ v_S]^T$, again using the Euclidian norm to quantify (dis)similarity between them. Note that we scale the subgraph counts so that we really measure a "density" of $s$ in $G$, given by:

$$\rho(G,s) = \binom{n}{k}^{-1} \sum_{\Phi:[k]\to[n]} \mathbb{1}[\forall i,j \in k : H(i,j) = G(\Phi(i), \Phi(j))],$$

where $n$ is the number of nodes in $G$, $k$ is the number of nodes in $s$, $\Phi$ is an injection from the first $k$ integers to the first $n$, and $\mathbb{1}$ is an indicator function which takes the value 1 when subgraph $s$ has been located in $G$. The subgraph enumeration procedure in the case of labeled nodes is illustrated in Fig. (5).

**An Unlabeled Graph Example: The Watts-Strogatz Model**

To validate the applicability of the above approach/graph similarity measure on graph objects, we used the Watts-Strogatz (WS) network generation model to construct a graph object dataset on which to apply DMAPS. For a fixed graph size, the WS model relies on two parameters to generate a so-called small-world network output.[19] The first parameter $p$ is initially used to generate a two-dimensional lattice where each vertex is connected to all of its neighbors situated a distance at most $p$ away. For each vertex in the graph, we choose the edge that connects it to its nearest neighbor and, with probability $r$, rewire this edge to connect with a vertex chosen uniformly at random from the graph. This procedure is repeated, each time considering the edge that connects the next closest neighbor to the vertex in question until all edges have been considered once, with no duplicate edges allowed. The resulting graph is considered the output of the model. The networks produced by this model exhibit various qualitatively different properties as its "generating parameters" vary. For $r = 1$ the model reduces to generating an Erdős–Rényi random graph $G(n,p)$: a graph of size $n$ for which the probability that any two vertices are connected is $p$. On the other hand, for $r = 0$ the graph remains a regular lattice, with no random changes in its topology. Lastly, for intermediate values of $r$ we get many local connections between adjacent nodes and few edges between far away nodes, which is a defining characteristic of small-world networks.

The WS algorithm was used to generate $n = 2000$ different small-world graphs, each with $n = 100$ vertices. For each such graph $G_i(p_i, r_i)$, we generated uniformly at random the two variables $r_i \sim \text{unif}[0,1]$ and $p_i \sim \text{unif}[0,1]$. We thus are confident that, by construction, this is a two-parameter set of graph data, parametrized by the generating parameters $p$ and $r$. The diffusion map embedding was then constructed by using the graph similarity measure defined above. It was found that the first two principal DMAP eigenvectors $\phi_2$ and $\phi_3$ were sufficient to represent the data. This is confirmed by the two dimensional nature of the $(\phi_2 - \phi_3)$ manifold. When considering the values of $p_i$ and $r_i$ of the various data points (the various graphs) that lie on this two-dimensional manifold, it can be observed that they vary in directions visually independent of each other. This strongly suggests that the transformation $f:(\phi_2, \phi_3) \rightarrow (r, p)$ has a nonsingular Jacobian matrix, which in turn implies that the transformation is bijective – the two variable pairs are one-to-one with each other, and they each constitute coordinates of the network dataset. This means that DMAPS discovered a reparameterization of the two variables $r$ and $p$, which in this case were known in advance to be (by construction) the variables that define this dataset. This serves as a validation for the DMAPS approach and the chosen graph similarity measure since, by only examining a dataset generated by the WS model, the technique was able to "learn" that only two features mattered, and that the two features were one-to-one with the construction parameters $(r, p)$ that here were *a priori* known.

**SIS Model Results**

In order to identify the coarse variables that parametrize the dynamics of the adaptive SIS model, diffusion maps were implemented on a graph dataset sampled from the SIS model evolution. This dataset was generated by systematically sampling graph objects from the SIS model simulation over time from various parameters/initial conditions, with each simulation leading to different long-term dynamical behavior. The principal directions, represented by the leading diffusion map coordinates, identify the important variables that define the model's evolution over time. Since the set of coarse variables $(i, l_{II}, l_{SS})$ are known to be central to the model's evolution, their relationship with the derived principal directions was investigated.[16]

More specifically, the system is known to undergo a Hopf bifurcation to periodic solutions as the parameter $p$ varies around $(p, r, w_0) \approx (0.00071, 0.0002, 0.06)$ and graph objects at parameter values around this bifurcation point were sampled to create a dataset of $N = 6000$ graphs $\{G_i\}_{i=1}^{N}$. After the graphs were sampled, the DMAPS procedure detailed above was applied, with our labeled graph metric used as a similarity measure. An analysis of the relationship between the diffusion map coordinates indicates a two-dimensional embedding in the first two principal directions $\phi_2, \phi_3$. Furthermore, an investigation of the relationship between $\phi_2$ and $\phi_3$ with other diffusion map coordinates demonstrates that no new direction is captured by higher order eigenvectors, something that strongly implies that the manifold on which the dataset lies is indeed two-dimensional.

Motivated by the evidence that the $(\phi_2, \phi_3)$ manifold fully captures all independent directions in the dataset, we look at the relationship between these two principal directions and the coarse variables $(i, l_{II}, l_{SS})$, known to encapsulate the long-term dynamics of this system. The relationship between $(\phi_2, \phi_3)$ and $(i, l_{II}, l_{SS})$ can be investigated by visually studying the embedding of the coarse variables in diffusion map space. Looking at the relationship between these three variables in our dataset, it can be noticed that they actually span two *(and not three different)* dimensions, which can be garnered by their two dimensional embedding in $\mathbb{R}^3$. Thus, we are actually looking for two principal directions, motivating the definition of $l_{SI} = L - (l_{II} + l_{SS})$, the total number of SI-links as a compound variable. This is done without introducing or removing any information from the system, as the total number of edges is constant throughout. We consider $\log(l_{SI})$ as a candidate macro-variable, since we are interested in finding a bijective relationship between $(\phi_2, \phi_3)$ and $(i, l_{SI})$.

By inspecting the relative directions of $i$ and $\log(l_{SI})$ in the two dimensional embedding of $(\phi_2, \phi_3)$, it becomes apparent that they are transverse to each other, with the former varying roughly from left to right and the latter from top to bottom. Such an observation is strong evidence that the Jacobian of the transformation $f:(\phi_2, \phi_3) \rightarrow (i, l_{SI})$ is nonsingular on this dataset, much in the same manner as for the Watts-Strogatz graph ensembles. Thus, we can conclude that the directions of change on the $\phi_2 - \phi_3$ manifold represented by changing $i$ and $l_{SI}$, respectively, are independent of each other, and that they are reparametrizations of the principal eigenvectors. Similar results are obtained if we consider $l_{SI}$ instead of $\log(l_{SI})$. These observations imply that the diffusion map technique has been successful in identifying, up to parameterization, the variables found in [16] as responsible for the long term dynamics of the model. Furthermore, we were able to confirm that the long-term dynamics of this model really depend on only two macro-variables, the total number of infected nodes $i$ and the total number of SI-links $l_{SI}$.

In addition, there was no need to take any *a priori* knowledge about the model's specifics into account when isolating the important variables. Instead, what was required was the development of a suitable similarity measure between labeled graphs, which can be generalized for use with other problems (models, datasets) that exhibit completely different dynamic behaviors. This generalization of feature extraction for high-dimensional systems can assist in developing a framework for isolating the coarse variables that define graph-based datasets without resorting to approaches that rely on intuition about the specific model.

**Conclusion**

The use of networks in the modeling of real-world phenomena is particularly suited to modeling the spread of an epidemic through a population connected through a network of geographic/social connections. In this paper, the suitability of data mining techniques for extracting the important macro-variables describing the evolution of network connectivity from an adaptive SIS model were investigated, based on two concepts. Firstly, a dissimilarity metric for graph objects needed to be constructed; secondly, it was used with the DMAPS nonlinear data-mining procedure. In particular, we needed to define a (dis)similarity measure between different *labeled* graphs, as they arise in the course of the epidemic model simulation. It should be noted that this method is generalizable to *any* labeled graph with distinct labels, and can be utilized regardless of the overarching process defining graph evolution. Future work in this area should investigate the suitability of other graph similarity measures for use within the DMAPS framework.

In addition, the long-term dynamics of a particular adaptive SIS model were explored, which allowed us to verify the suitability of the constructed metric for use with graph object datasets. This yields two interconnected results. Firstly, the DMAP procedure was able to demonstrate that the system is inherently macroscopically two dimensional, with the two independent directions in the dataset being represented by the first two nontrivial Diffusion Map eigenvectors. This result is consistent with previous knowledge about the dimensionality of the model's dynamics, and is strong evidence that the DMAP framework, using the similarity measure we constructed, can be readily applied to graph object data sets to extract meaningful reduced parametrizations of the underlying behavior.

Furthermore, we were also able to link the principal diffusion map coordinates with previously known coarse variables capable of describing this system. By inspection of the embeddings in diffusion space, it was verified that a bijection exists between two of these coarse variables and the two leading diffusion map coordinates. This means that the DMAP was not only able to learn the inherent dimensionality of the dataset, but that it was also able to extract a reparameterization of variables known to fully specify this model; what is unfortunate, is that the variables of this reparametrization have no obvious unique, easily explainable physical meaning.

This modeling exercise clearly shows that modern data mining techniques for data in the form of high-dimensional evolving vectors can be extended to data in the form of large evolving graphs (labeled or unlabeled). This holds promise for the analysis of data from epidemics on realistic adaptive networks, and for general adaptive network evolution problems. What is more important and more promising, however, is that these data-based coarse descriptors can be used, in an equation-free framework, to implement accurate reduced model computations for the epidemic dynamics, in which the variables used to describe the systems-level network behavior are the ones obtained from data mining. This data-driven model reduction approach can be introduced as a "wrapper algorithm" around brief bursts of detailed, fine scale simulation; we believe that the approach holds real promise in enabling systems level analysis, simulation and control of detailed, realistic epidemic dynamics.

**Acknowledgements**. This work was partially supported by the US National Science Foundation (CDS&E program).

**Disclosure of Potential Conflicts of Interest.** No potential conflicts of interest were disclosed.

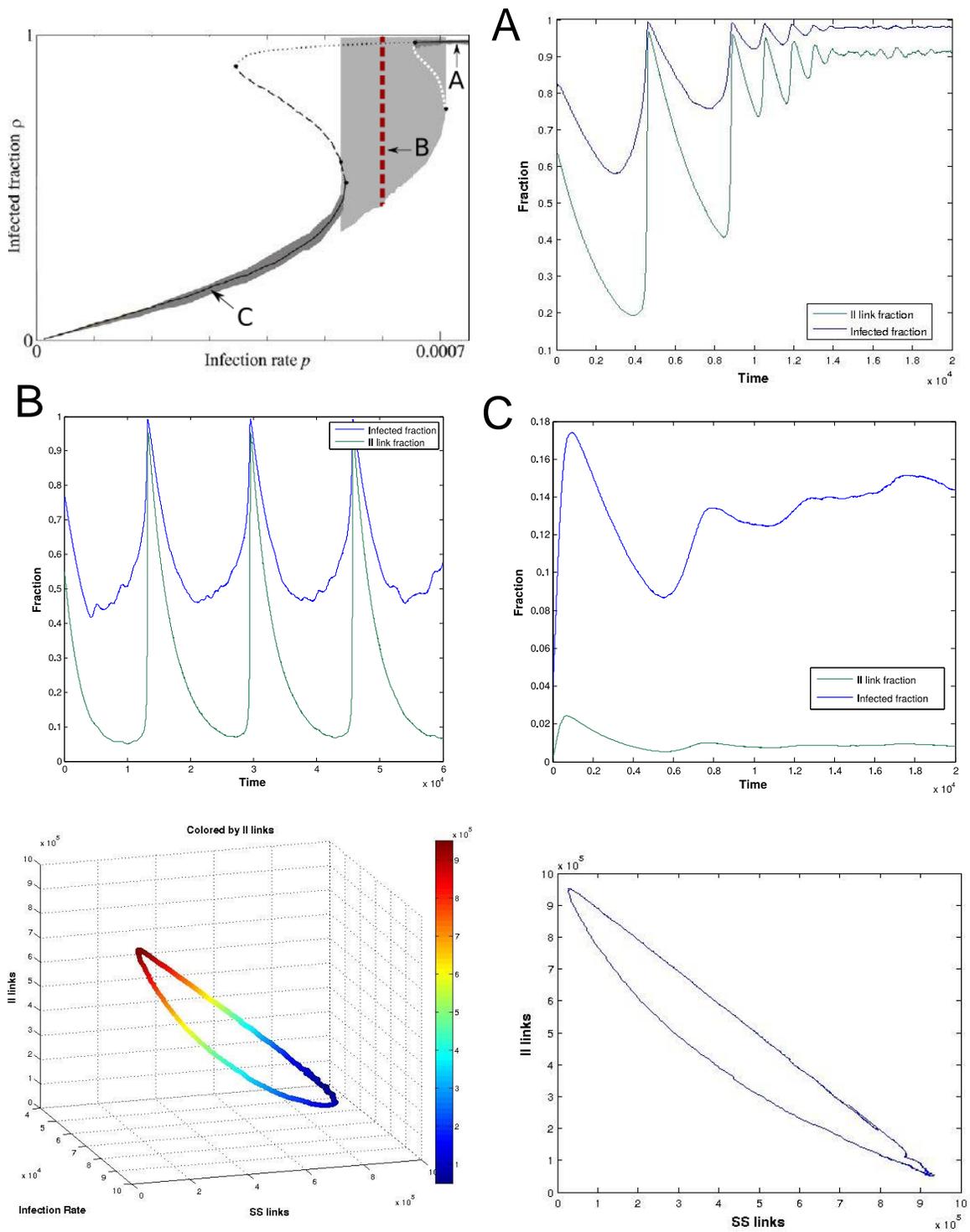

Figure 1: (Top left) Model bifurcation diagrams wrt. the infection rate parameter p (reprinted with permission).[16] (A) The system evolves to a stable stationary state for $p = 0.00073$. (B) Oscillatory behavior indicating a (coarse) limit cycle at $p = 0.0006$. (C) Stable stationary state for $p = 0.0003$. Bottom graphs indicate the relationship between $i$, $l_{SS}$, and $l_{II}$ over the course of one complete oscillation for $p = 0.0006$. Model parameters: $(r, w_0, N, L) = (0.0002, 0.03, 10^5, 10^6)$.

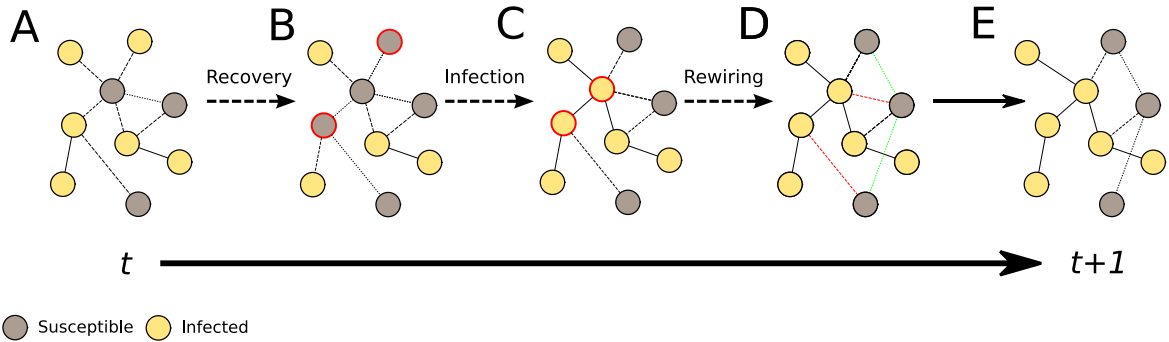

Figure 2: Schematic of the adaptive SIS model evolution: the three substeps constituting one SIS evolution timestep. (A) The initial graph at time $t$. (B) Each infected node recovers with probability $r$, becoming susceptible. (C) The disease spreads along SI-links with probability $p$, infecting susceptible nodes. (D) With probability $w$, each SI-link is broken and a new SS-link is created. (E) The final graph at $t+1$. Broken links and nodes that change status between steps are colored in red, while rewired links are colored in green.

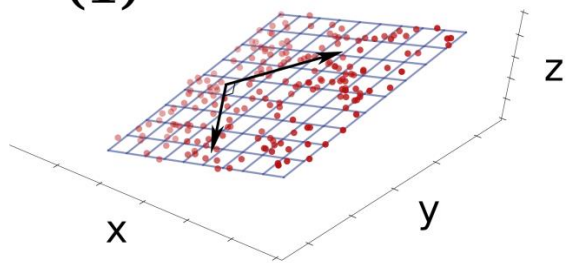

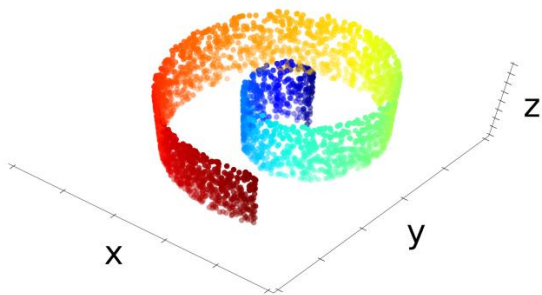 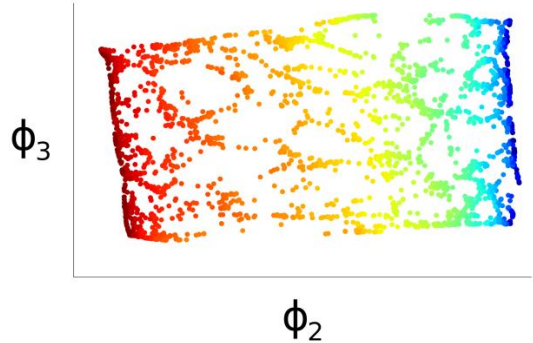

Figure 3: (1) PCA uncovers the linear relationship (two-dimensional blue grid spanned by two black arrows) within a noisy dataset of three-dimensional points lying approximately on a plane (red dots). (2a) Three-dimensional data on a *nonlinear, curved surface*. Color denotes arclength. (2b) DMAPS embedding uncovering a two-dimensional parameterization of the dataset. Color denotes arclength from (2a).

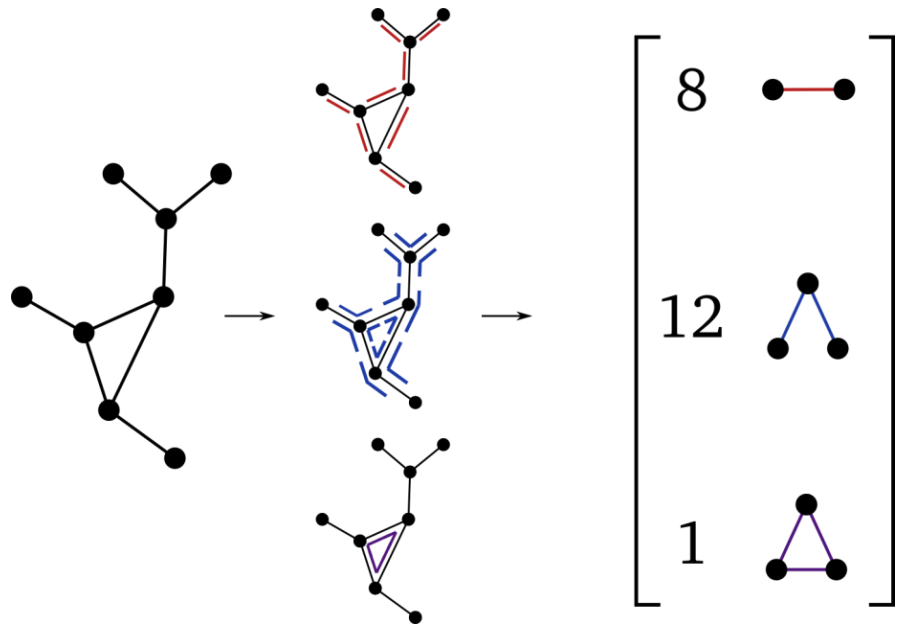

Figure 4: Illustration of the subgraph-enumeration process with an unlabeled input graph.

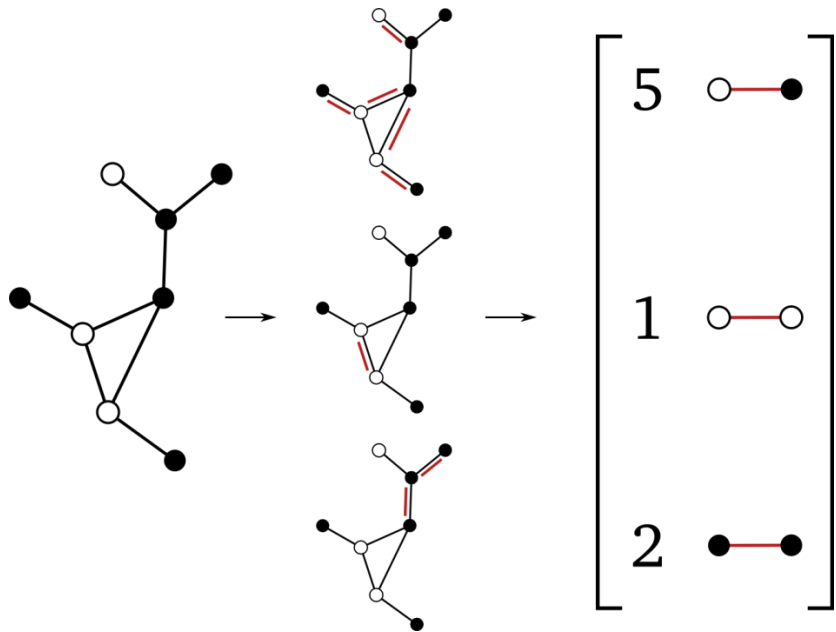

Figure 5: Illustration of the subgraph-enumeration process with a labeled input graph. In this case, we must discriminate between differently labeled subgraphs.

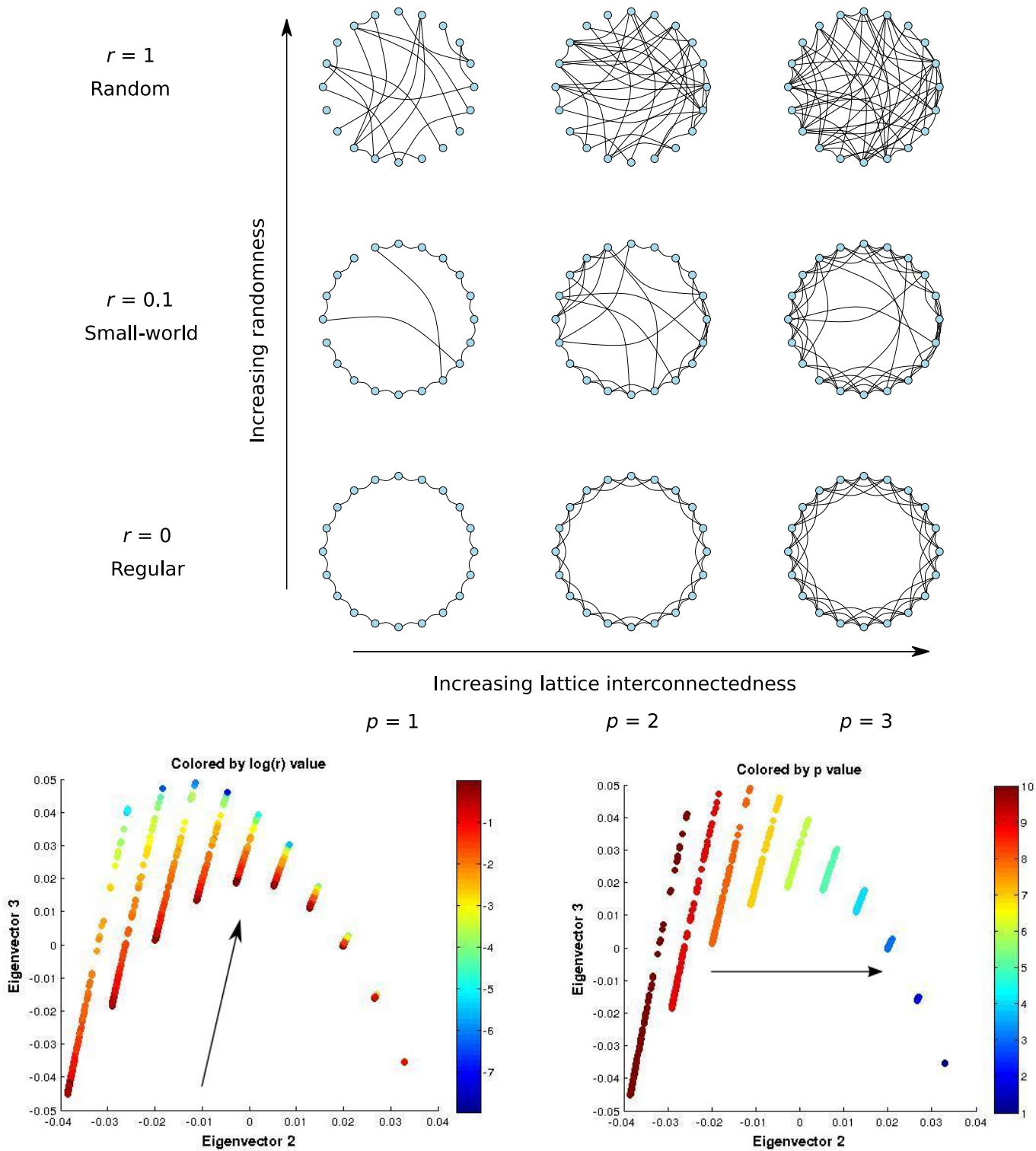

Figure 6: The Watts-Strogatz Model and its construction parameters: The value of $r$ denotes the probability of long-distance rewiring with $r = 0$ denoting a regular lattice, $r < 1$ a small-world graph, and $r = 1$ an Erdős–Rényi random graph. The value of $p$ quantifies how interconnected the initial lattice is, being the number of neighboring nodes each node connects to. *Below:* Second versus third eigenvector ($\Phi_2 - \Phi_3$) colored by $\log(r)$ (*left*) and $p$ (*right*). The visual linear independence of the directions of change of $\log(r)$ and $p$ indicate that the ($\Phi_2$, $\Phi_3$) coordinates form a bijection with (a reparametrization of) the construction parameters ($r$, $p$). An ε of 0.1 was used in our DMAPS computations.

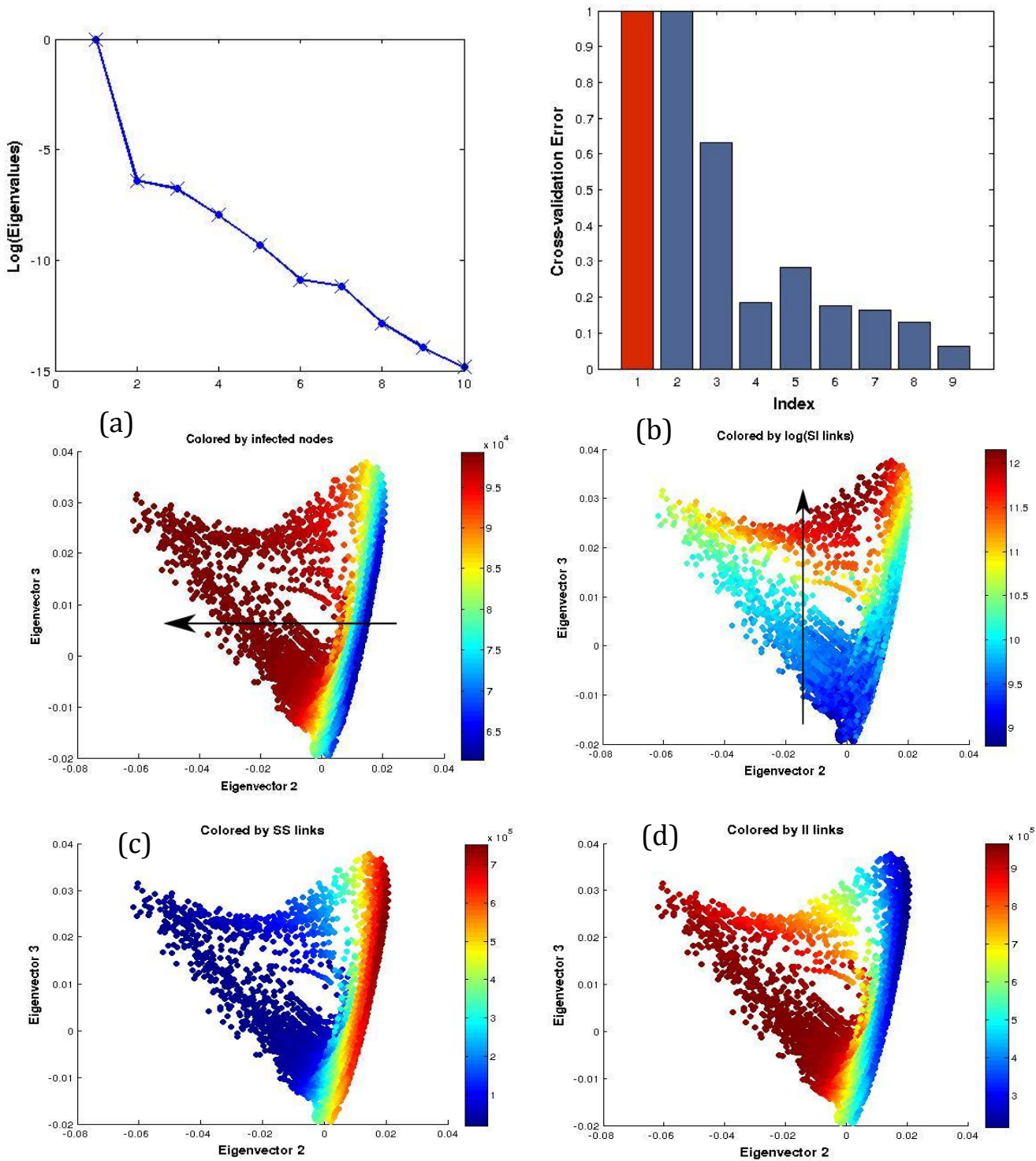

Figure 7: Coarse variable detection: Top Left: The leading eigenvalues of the random walk matrix are plotted. Top Right: Computation a criterion suggesting that the first two eigendirections suffice (the first is a trivial one, and the third is much less important than the first two. The *x* and *y* coordinates of the middle and bottom row figures indicate the components of each datum, representing a graph, in the second and third eigenvectors of the random walk matrix. Each point is also colored by the number of (a) infected nodes *i*, (b) log(SI-links), (c) SS-links, and (d) II-links found in the corresponding graph. A comparison between (a) and (b) indicates a linear independence relationship between *i* and $\log(l_{SI})$. An ε of 70 was used.

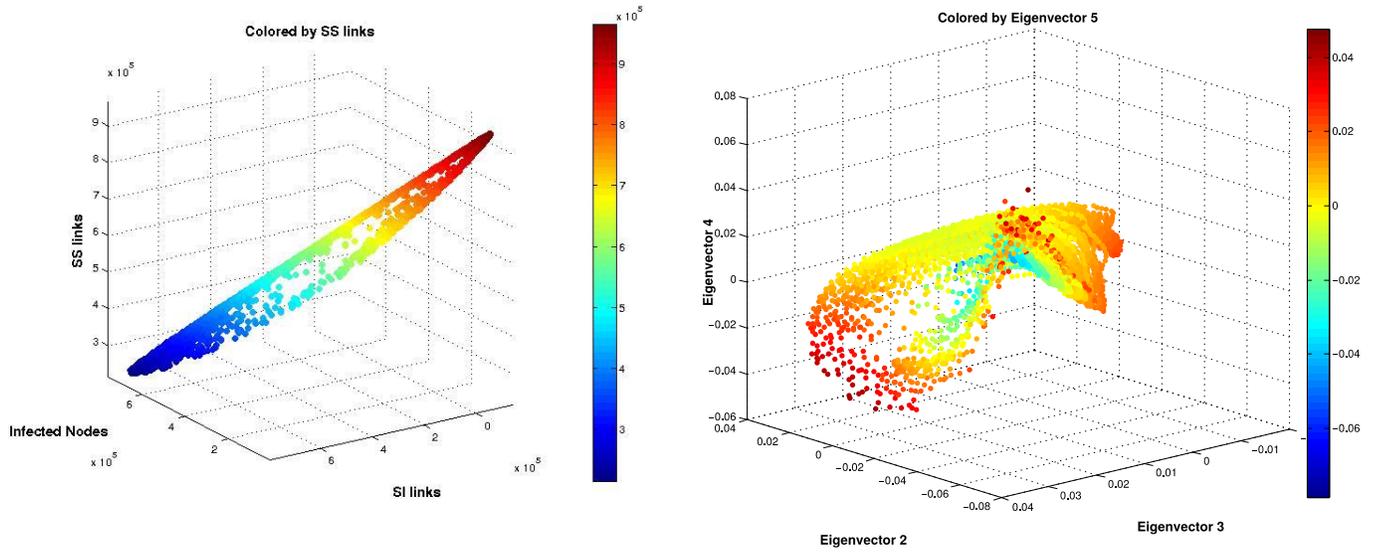

Figure 8: Coarse Variable Dependencies: An illustration of the relationship between $l_{SS}$, $l_{SI}$, and $i$ in the diffusion map dataset. The two-dimensional nature of the manifold indicates that the number of SS links can be thought of as a function, on the data, of $i$ and $l_{SI}$, and is thus not necessary as an independent macro-variable.